\documentclass[copyright]{eptcs}
\providecommand{\event}{TFPIE} 
\usepackage{breakurl}             

\title{Mathematics Is Imprecise}
\author{Prabhakar Ragde
\institute{Cheriton School of Computer Science\\
University of Waterloo\\
Waterloo, Ontario, Canada}
\email{plragde@uwaterloo.ca}
}
\def\titlerunning{Math Is Imprecise}
\def\authorrunning{P.~Ragde}
\begin{document}
\maketitle

\begin{abstract}
  We commonly think of mathematics as bringing precision to
  application domains, but its relationship with computer science is
  more complex.  This experience report on the use of Racket and
  Haskell to teach a required first university CS course to students
  with very good mathematical skills focusses on the ways that
  programming forces one to get the details right, with consequent
  benefits in the mathematical domain. Conversely, imprecision in
  mathematical abstractions and notation can work to the benefit of
  beginning programmers, if handled carefully.
\end{abstract}

\section{Introduction}

Mathematics is often used to quantify and model what would otherwise
be poorly-understood phenomena. However, as an activity carried out by
humans for humans, it can and does take advantage of imprecision:
using ambiguous notation, omitting cases that are ``similar,'' and
eliding details. The machines that mediate activity by humans for
humans in computer science introduce an element of forced
precision. The thesis of this paper is that pedagogical attention to
this relationship can enhance learning in both disciplines, by
introducing more precision to mathematics, and by careful use of
imprecision in computer science.

The University of Waterloo has the world's largest Faculty of
Mathematics, with six departments (including a School of Computer
Science), over 200 faculty members, and about 1400 undergraduate
students entering each year. These students are required to take two
CS courses, and they have a choice of three streams. Two are aimed at
majors and non-majors respectively; the third is aimed at students
with high mathematical aptitude. A similar high-aptitude stream has
existed for the two required math sequences (Calculus and Algebra) for
decades, but the CS advanced stream is relatively recent, starting
with a single accelerated course in 2008 and moving to a two-course
sequence in 2011-2012.

The CS advanced stream currently has a target of 50-75 students per
year.  Admission is by instructor consent, or by scoring sufficiently
high on math or programming contests at the senior high-school
level. Consequently, a significant fraction (sometimes more than half)
of the students taking the advanced stream are not CS majors (and many
who are will take a second major in one of the other Math
departments).  Some students have considerable experience in
imperative programming, while others have no programming experience at
all.  Functional programming, with its low barriers to entry and its
elegant abstractions, is well-suited to provide the right sort of
challenges for such a diverse population.

Our major and non-major streams use Racket \cite{R} exclusively in the
first course, with the ``How To Design Programs'' (HtDP) textbook
\cite{HtDP} and the Program By Design (PBD) methodology \cite{PBD}.
(The second courses make a gradual transition to C for majors and
Python for non-majors.) Because of the difficulty of assessing
placement (many non-majors would be better off with the moderate
challenge of the major course, and the advanced course also draws from
both groups) and consequent student migration between
streams, the advanced stream cannot stray too far from this model,
but some deviation is possible. The rest of the curriculum ignores
functional programming, so upward compatibility is not an issue.

There are thus some major similarities among the first courses in all
three streams, and indeed with courses on functional programming using
other languages and textbooks: starting with the manipulation of
numbers and structures with a fixed number of fields, introducing
recursion with lists, and continuing with trees. PBD emphasizes
data-directed design, and the use of examples and tests to guide code
development. 

In the remainder of this paper, I will describe some unusual choices
that I made in the design of the first advanced course, some
techniques that seemed to find favour with students, and some issues
that remain to be overcome.

\section{The roles of Racket and Haskell}

Among institutions using a functional-first approach, Haskell \cite{H}
is a popular choice. Haskell is an elegant and highly-expressive
language, and its proximity to mathematics would make it a natural
choice for students in the advanced stream. Thus the reader may be
surprised at the choice I made in the first advanced course: while the
first set of lectures uses Haskell exclusively, and students see it
throughout the advanced course, all of their assignment programming is
done in Racket. Haskell is used as functional pseudocode.

Conventional pseudocode, at its best, resembles untyped Pascal:
imperative, with loops manipulating arrays and pointers. In
comparison, code written in a functional language is transparent
enough that it often serves the same purpose. However, there are
degrees of transparency, and some functional languages are more
readable than others. Haskell, with patterns in function definitions
and local bindings, and infix notation, is rich in expressivity, and
it is highly readable as long as care is taken to not make it too
terse (at least on early exposure).

However, students actually programming in Haskell (as opposed to just
reading it for comprehension) have to learn about operator
precedence, and have to learn the pattern language. Mistakes in these
areas often manifest themselves as type errors, aggravated by type
inference making interpretations that the student does not yet know
enough to deliberately intend or avoid, and compiler errors designed
to inform the expert. Well-written Haskell code is a joy to read;
poorly-written, incorrect Haskell code can be a nightmare for the
beginner to fix.

Racket's uniform, parenthesized syntax (inherited from Lisp and
Scheme) is by contrast relatively straightforward; the teaching
language subsets implemented by the DrRacket IDE limit student errors
that produce ``meaningful nonsense''; and testing is lightweight,
facilitating adherence to the PBD methodology. Seeing two languages
from the beginning lets students distinguish between concepts and
surface syntax (in effect providing them with a basis for
generalization), while programming in just one minimizes operational
confusion.  When I introduce more advanced features available in full
Racket (such as pattern matching and macros), students can appreciate
them (with the foreshadowing provided by Haskell) and put them to use
immediately.

Following Hutton, who in his textbook ``Programming In Haskell''~\cite{Hu}
does not even mention lazy evaluation until the penultimate chapter, I
am vague about the computational model of Haskell at the
beginning. But a precise computational model is important in
debugging, and the simplified reduction semantics that HtDP presents
is quite useful, especially combined with the DrRacket tool (the
Stepper) that illustrates it on student code.

\clearpage

In fact, though the code I show is legal Haskell (with a few elisions,
such as the use of \verb|deriving Show| or type signatures necessary
to assuage the compiler), as pseudocode it should perhaps be called
``Raskell,'' because, in early computational traces and later analysis
of running time, I assume strict (not lazy) semantics corresponding to
those of Racket.

\section{Computation and proof}

Here is the first program that the students see.

\begin{verbatim}
data Nat = Z | S Nat
plus x Z     = x                
plus x (S y) = S (plus x y)
\end{verbatim}

Peano arithmetic is not normally treated in a first course on
computing, though it may show up in a later course on formal logic or
a deep enough treatment of Haskell to show its utility in advanced
notions of types. One reason to introduce it here is that the Algebra
course my students are taking simultaneously is not linear algebra,
but ``classical algebra'', which uses elementary number theory to
illustrate the process of doing mathematics. However, that course
assumes the properties of integers as a ring and rational numbers as a
field (without using those terms), as does every math course before a
formal treatment of groups, rings, and fields. This gives us an
opportunity to show that computers cannot just assume these operations
exist, but must implement them.

HtDP distinguishes three kinds of recursion: structural recursion,
where the structure of the code mirrors a recursive data definition
(as above);
accumulative recursion, where structural handling of one or more
parameters is augmented by allowing other parameters to accumulate
information from earlier in the computation (illustrated below); 
and generative recursion,
where the arguments in a recursive application are ``generated'' from
the data (early examples include GCD and Quicksort). 

A computational treatment of Peano arithmetic respects this hierarchy
(the code above is structurally recursive) while immediately serving
notice that mathematical assumptions will be challenged and details
are important. Being precise about addition, an activity students have
carried out almost as long as they can remember, but which they likely
have not examined carefully, gives a fresh perspective on
mathematics. This approach also permits me to
address in a timely fashion the notion of proofs and their importance
to computer science.

The first proof they see is an example of classic
$\forall$-introduction, where a free variable in a proved statement
can be quantified. Here is a proof of 
``\texttt{add x (S (S Z)) = S (S x)}''.

\begin{verbatim}
add x (S (S Z)) = S (add x (S Z))
                = S (S (add x Z))
                = S (S x)
\end{verbatim}

We can now conclude 
``For all \texttt{Nats x}, \texttt{add x (S (S Z)) = S (S x)}''.
I describe this to the students as ``the
anonymous method''; the emphasis here is another example of greater
precision in mathematics than is typical at this level, where implicit
for-all quantification is a source of much confusion. (Note the
computational model here, a restricted form of equational reasoning
where the clauses of the function definition are treated as rewriting
rules. This meshes quite well with the reduction semantics given for
Racket.) 

The anonymous method is inadequate for a proper exploration of proof,
even at this point. Attempts to prove, for example, commutativity or
associativity (other concepts they have taken for granted) founder. An
even simpler example is 
``For all \texttt{Nats x}, \texttt{add Z x = x}''. We can prove this
for small examples, such as \texttt{x = S (S (S Z))}:

\begin{verbatim}
add Z (S (S (S Z))) = S (add Z (S (S Z)))
                    = S (S (add Z (S Z)))
                    = S (S (S (add Z Z)))
                    = S (S (S Z))
\end{verbatim}

At this point the student can see the proof for the case 
\texttt{x = S (S Z)}, on the right hand side if one layer
of \texttt{S} is stripped away. In this way, we arrive at the need for
and justification of structural induction on our definition of \texttt{Nat}.
They see induction in their Algebra sequence (immediately in
the advanced stream, after a few weeks in the regular stream) but it
is not applied to ``fundamental'' properties of arithmetic, which are
taken for granted.

This approach falls short of full formalism, either through a proof
assistant such as Coq or ACL, or through a classic presentation of
Peano arithmetic in the context of formal logic, either of which would
be overkill for an introductory course. Instead, it uses
computer science and mathematics together to yield more insight than
traditional pedagogical approaches at this level in either discipline.

Discussing proofs by induction also reinforces the idea that
structural recursion, should it work for the problem at hand, 
is a preferable approach, as it is easier
to reason about, even informally. We look at a
non-structurally-recursive version of addition:
\begin{verbatim}
data Nat = Z | S Nat
add x Z     = x                
add x (S y) = add (S x) y
\end{verbatim}
This function uses accumulative recursion (the first parameter is an
accumulator), and it is harder to prove properties such as the one above,
commutativity, or associativity. In fact, the easiest way to do this
is to prove that \texttt{add} is equivalent to \texttt{plus}, and then
prove the properties for \texttt{plus}.

Surprisingly, this situation carries over into many early uses of
accumulative recursion, such as to add up or reverse a list. An
accumulator resembles a loop variable, and the correspondence is
direct in the case of tail recursion. The conventional approach to
proving correctness is to specify a loop invariant that is then proved
by induction on the number of iterations (or, in the functional case,
the number of times the recursive function is applied). But it turns
out that a direct proof (by structural induction) that the
accumulatively-recursive function was equivalent to the
structurally-recursive version is, in many cases, easier and
cleaner. The reason is that many of the standard proofs of loop
invariants involve definitions that use notation (such as $\Sigma$ for
addition) whose properties themselves require recursive definitions
and proofs.

As an example, consider adding up a list.
\begin{verbatim}
sumh [] acc     = acc
sumh (x:xs) acc = sumh xs (x+acc)
sumlist2 xs = sumh xs 0
\end{verbatim}
An informal proof of correctness of \texttt{sumlist2}, based on Hoare
logic, would use an invariant such as ``In every application of the
form \texttt{sumh ys acc}, the sum of the whole list is equal to
\texttt{acc} plus the sum of the \texttt{ys}.''  But there really is
no better formalization of ``the sum of'' in this statement than the
structurally recursive definition of \texttt{sumlist}:
\begin{verbatim}
sumlist []     = 0
sumlist (x:xs) = x + sumlist xs
\end{verbatim}
At which point it is easier and more straightforward
to prove ``For all \texttt{xs}, for all \texttt{acc},
\texttt{sumh xs acc = acc + sumlist xs}'' by structural induction on
\texttt{xs}. We arrive at this only by trying to prove the more
obvious statement ``For all \texttt{xs},
\texttt{sumh xs 0 = sumlist xs}'' and failing, because the inductive
hypothesis is not strong enough. The difficulty of finding an
appropriate generalization to capture the role of the accumulator
(which gets harder with more complex code)
underlines the difficulty of understanding and informally justifying
code that uses an accumulator.

The strong connection between structural recursion and structural
induction makes it possible to discuss rigourous proofs of correctness
in a way that is not overwhelming (as it typically is for Hoare
logic), and this extends to most uses of accumulative
recursion. Traditional invariants are easier to work with in the
absence of mutation than if it is present, but they still require more
work than the direct approach of structural induction. Strong
induction, or induction on time or number of recursive applications,
can thus be deferred until generative recursion is taught.

\section{Analyzing efficiency}

A traditional CS1-CS2 approach defers discussion of algorithm analysis
and order notation to the second course, leaving the first one to
concentrate on the low-level mechanics of programming. However,
efficiency influences not only the design of imperative languages, but
the ways in which elementary programming techniques are
taught. Efficiency is also the elephant in the room in a
functional-first approach, though the source of the problem is
different. A structurally-recursive computation where it is natural to
repeat a subexpression involving a recursive application (for example,
finding the maximum of a nonempty list) leads to an exponential-time
implementation, with noticeable slowdown even on relatively small
instances. The fix (moving code with repeated subexpressions to a
helper function) is awkward unless local variables are
prematurely introduced, and even then, the motivation has to be
acknowledged. Accumulative recursion is also primarily motivated by
efficiency.

Our major stream also postpones order notation to the second course,
while reluctantly acknowledging the elephant where necessary. The
advanced stream, however, introduces order notation early. An
intuitive illustration of time and space complexity is easy with our
first example of unary numbers, as it is clear from a few traces that
our representation takes up a lot of room and computation with it is
slower than by hand. We more carefully exercise these ideas by moving
at this point into a sequence of lectures on
representing sets of integers by both unordered and ordered lists.

Order notation shares pedagogical pitfalls with another topic
commonly introduced in first year, limits in calculus. Both concepts
have precise definitions involving nested, alternating quantifiers,
but students are encouraged to manipulate them intuitively in a
quasi-algebraic fashion. A typical early assignment involves questions
like ``Prove that $6n^2 - 9n -7$ is $O(n^2)$.'' As with epsilon-delta
proofs, not only do weaker students turn the crank on the form without
much understanding, but questions like this have little to do with
subsequent use of the ideas. The situation is worse with order notation
(more quantifiers, discrete domains that are difficult to visualize).

The analysis of imperative programs at the first-year level is little
more than adding running times for sequential blocks and multiplying
for loop repetitions; in other words, it is compositional based on
program structure. The obvious approach for recursive functions
involves recurrences. But solving recurrences is not easy, even with
standard practices such as omitting inconvenient floors and ceilings,
and setting up recurrences is not straightforward, either. I have
found that a compositional approach works for many recursive
functions encountered in this course, with the aid of a table.

The tabular method works for functions that use structural or
accumulative recursion, as long as the recursive application is done
at most once on each ``piece'' of the argument corresponding to a
self-referential part of the data definition. For lists, this means the
``rest'' of the list; for binary trees, this means the two subtrees. All the
functions they need to write in early treatment of lists and binary trees are
structurally or accumulatively recursive.

Racket functions consuming data of these forms consist of a
\texttt{cond} at the top level, and the table has one row for each
question-answer pair (equivalently, for each pattern plus guard in a
Haskell multipart definition). The row contains entries for the number
of times the question is asked (as a function of the ``size'' of the
argument), the cost of asking the question (nearly always constant),
the number of times the answer is evaluated, and the cost of
evaluating the answer (apart from recursive applications). These are
multiplied in pairs and added to give the cost of the row, and then
these costs are added up over all rows. Here is how the table might
look for \texttt{sumlist} (where $n$ is the length of the list
argument):

\smallskip
\begin{tabular}{c|c|c|c|c|c}
Row&\#Q&time Q&\#A&time A&total\\ \hline
1 &$n+1$&$O(1)$&$1$&$O(1)$&$O(n)$\\
2 &$n$  &$O(1)$&$n$&$O(1)$&$O(n)$\\ \hline
  &     &      &   &      &$O(n)$\\
\end{tabular}

\smallskip

For a function with more than two cases, we typically cannot be so
precise about the number of questions and answers. Order notation once
again comes to the rescue.

\begin{verbatim}
filter p [] = []
filter p (x:xs) 
 | p x       = x : filter p xs
 | otherwise = filter p xs
\end{verbatim}

Here is the tabular analysis of the running time of \texttt{filter}
on a list of length $n$.

\smallskip
\begin{tabular}{c|c|c|c|c|c}
Row&\#Q&time Q&\#A&time A&total\\ \hline
1 &$n+1$ &$O(1)$&$1$&$O(1)$&$O(n)$\\
2 &$O(n)$&$O(1)$&$O(n)$&$O(1)$&$O(n)$\\
3 &$O(n)$&$O(1)$&$O(n)$&$O(1)$&$O(n)$\\ \hline
  &     &      &   &      &$O(n)$\\
\end{tabular}

\smallskip
This approach does not entirely avoid recurrences, which are necessary
to explain, for example, the exponential-time behaviour of na\"ive
list-maximum, but it limits their use.

Here we are using the imprecision of order notation in two different
ways. The loss of information about the exact running time streamlines
the analysis by not carrying along irrelevant detail. We are also
working with an intuitive or fuzzy understanding in the heads of
students as to the meaning of an order-notation assertion (it is still
easy, when using the tabular method, to erase the distinction between
the $n^2$ appearing in a table entry and the actual running time that
it bounds, qualified by the appropriate constants). While this
can lead them into difficulty in more pathological situations, it
suffices for the kind of analyses necessary at the first-year level.

\clearpage

\section{Efficient representations of integers}

The approach I take to the efficient representation of integers starts
by arguing that the problem with unary arithmetic stems from the use
of a single data constructor with interpretation \texttt{S}:$n\mapsto
n+1$. Using two data constructors, we must decide on interpretations.
\begin{verbatim}
data Nat = Z | A Nat | B Nat
\end{verbatim}
Effective decoding requires that the range of the two interpretations
partition the positive integers. ``Dealing out'' the positive integers
suggests an odd-even split, with interpretations \texttt{A}: $n\mapsto
2n$ and \texttt{B}: $n\mapsto 2n+1$. This leads to a form of binary representation
(with the rightmost bit outermost), with unique representation
enforced by a rule that \texttt{A} should not be applied to
\texttt{Z} (corresponding to the omission of leading zeroes). 
The interpretation easily yields a structurally recursive
\texttt{fromNat} to convert to standard numeric representation, and
its inverse \texttt{toNat}.

\begin{verbatim}
toNat 0 = Z

toNat 1 = B Z

toNat 2 = A (B Z)

toNat 3 = B (B Z)

toNat 4 = A (A (B Z))       
\end{verbatim}

We cover addition and multiplication in the new representation, and
analyze them. This leads to an interesting side effect. Mutual
recursion is introduced in HtDP in the context of trees of arbitrary
fan-out. But it arises naturally with the linear structures used
here. 

A first attempt at addition might look like this:
\begin{verbatim}
add x Z = x
add Z y = y

add (A x) (A y) = A (add x y)
add (A x) (B y) = B (add x y)
add (B x) (A y) = B (add x y)
add (B x) (B y) = A (add1 (add x y))

add1 Z = B Z
add1 (A x) = B x
add1 (B x) = A (add1 x)
\end{verbatim}

A na\"ive analysis of \texttt{add} first analyzes \texttt{add1}, which
takes $O(s)$ time on a number of size $s$ (number of data constructors
used in the representation). Then \texttt{add} takes time $O(m^2)$,
where $m$ is the size of the larger argument. However, this analysis
is too pessimistic. \texttt{add} actually takes time $O(m)$, since the
total work done by all applications of \texttt{add1} is $O(m)$, not
just one application. This is because the recursion in \texttt{add1}
stops when an \texttt{A} is encountered, but the result of applying
\texttt{add1} in \texttt{add} is wrapped in an \texttt{A}.

But this argument is subtle and difficult to comprehend. It is better
to replace the last line in the definition of \texttt{add} with an
application of an ``add plus one'' function.
\begin{verbatim}
add (B x) (B y) = A (addp x y)
\end{verbatim}
We then develop \texttt{addp}, which has a similar structure to
\texttt{add}, and recursively applies \texttt{add}. It is now easy to
see that \texttt{add} has running time linear in the size of the
representation, because it (or \texttt{addp}) reduces the size of the
arguments at each step. 

Another surprising benefit of this approach is that we can easily
represent negative numbers simply by introducing the new nullary
constructor \texttt{N}, representing $-1$. The interpretations of
\texttt{A} and \texttt{B} remain the same, as do the representations
of positive numbers; we add the rule that
\texttt{B} cannot be applied to \texttt{N}. The resulting
representation of integers is isomorphic to two's complement
notation.

\begin{verbatim}
toInts (-1) = N

toInts (-2) = A N

toInts (-3) = B (A N)

toInts (-4) = A (A N)

toInts (-5) = B (B (A N))
\end{verbatim}

The more traditional representation of two's complement can be seen by
reading right-to-left and making the following substitutions: 0 for
\texttt{A}, 1 for \texttt{B}, the left-infinite sequence of 0's for
\texttt{Z}, and the left-infinite sequence of 1's for \texttt{N}.

\begin{verbatim}
 3 = ...011
 2 = ...010
 1 = ...01
 0 = ...0
-1 = ...11
-2 = ...10
-3 = ...101
-4 = ...100
-5 = ...1011
\end{verbatim}

When we work out addition for the extended representation, we discover
that the existing rules for \texttt{add} stay the same, and the new
ones involving \texttt{N} are easy to work out.  Two's complement
notation is normally mystifying to second-year students taking a
computer architecture course, because it is presented as a polished
technique that ``just works'' (that is, reuse of the logic for
unsigned binary addition, with just a little added circuitry). Here we
have not only a clear explanation of how it works, but good motivation
for the development. The internal representation of numbers in both
Racket and Haskell is no longer magic.

The savings in space and time are intuitive, but when we quantify
them, we can introduce and solve exactly the recurrence
relating a natural number $n$ to the size of its representation, which
is an effective introduction of logarithms to the base 2 that does not
duck issues of discretization.

\clearpage

\section{Efficient representations of sequences}

Trees are often introduced to mirror structure in data: in HtDP, using
family trees, and in our major sequence, using phylogeny trees. An
important insight is that introducing tree structure to data not
obviously structured in this fashion can yield improvements in
efficiency. Unfortunately, the example usually chosen to illustrate
this, binary search trees, is not effective at the first-year
level. The simplest algorithms are elegant but degenerate to lists in
the worst case; there are many versions of balanced search trees, but
the invariants are complex and the code lengthy, particularly for
deletion. As a result, first-year students only see artificial
examples of balanced trees, such as the ones that can be built from an
already-sorted sequence of keys. 

Of course, this material is important, and we do treat it. But the
first example should be a success. The first introduction of a tree
structure to data for purposes of efficiency should result in a
quantifiable improvement, one that is not deferred to an intermediate
data structures course in second year or later.

The treatment of natural numbers in the previous section provides a
path to an effective introduction of logarithmic-height binary
trees. Consider the problem of representing a sequence of elements so
as to allow efficient access to the $i$th element. A list can be
viewed as being indexed in unary, with the element of index \texttt{Z}
stored at the head and the tail containing the sequence of elements of
index \texttt{S x}, stored in the same fashion but with the common
\texttt{S} removed from all indices. The reason it takes $O(i)$ time
to access the $i$th element of a list is similar to the reason it
takes $O(i)$ time to add the unary representation of $i$ to another number.

Binary representation of numbers suggests storing two subsequences
instead of one: the sequence of elements of index \texttt{A x}, and
the sequence of elements of index \texttt{B x}. This leads to the idea
of a binary tree where an element of index \texttt{A x} is accessed by
looking for the element of index \texttt{x} in the left
(``\texttt{A}'') subtree, and an element of index \texttt{B x} is
accessed by looking for the element of index \texttt{x} in the right
(``\texttt{B}'') subtree. This is just an odd-even test, as used in
\texttt{toNat}, and the reader will recognize the concept of a binary trie.

But there is a problem in this particular application, stemming from
the lack of unique representation and our ad-hoc rule to get around
it. Not all sequences of \texttt{A}'s and \texttt{B}'s are possible,
since \texttt{A} cannot be applied to \texttt{Z}. This means that
roughly half the nodes (every left child) have no element stored at
them, since that element would have an index ending with \texttt{A
  Z}. We can avoid this problem by starting the indexing at 1, or,
equivalently, retaining indexing starting at 0 but ``shifting'' to
1-based before applying/removing \texttt{A} or \texttt{B} and then
shifting back. In other words, we can replace the \texttt{A-B}
representation with a \texttt{C-D} representation, with interpretation
\texttt{C}$(n)$ = \texttt{A}$(n+1)-1$ and \texttt{D}$(n)$ =
\texttt{B}$(n+1)-1$.

This results in the interpretation \texttt{C}: $n\mapsto 2n+1$ and
\texttt{D}: $n \mapsto 2n+2$. Conversion between the new \texttt{C-D}
representation and built-in integers is as simple as with the old
\texttt{A-B} representation. The new representation is naturally unique
(without the need for extra rules), and all sequences are possible, so
there are no empty nodes in the tree with ``\texttt{C}'' left subtrees
and ``\texttt{D}'' right subtrees.  It is easy to show (again, by
solving a recurrence exactly) that the tree has depth logarithmic in
the total number of elements. Furthermore, not only does access to the
$i$th element takes time $O(\log i)$ by means of very simple
purely-functional code, but standard list operations (cons, first,
rest) take logarithmic time in the length of the sequence. We have
rederived the data structure known as a Braun tree \cite{BR}.
The code for deletion (rest) is no more complicated than the code for
addition; indeed, there is a pleasant symmetry.

Our attention to mathematical detail in the treatment of natural
numbers has paid off with an unexpected and fruitful connection to
purely-functional data structures.
We see that a more mathematical treatment of fundamentals is not in conflict
with core computer science content; on the contrary, it supports the content
and increases accessibility by providing sensible explanations for choices.

\section{Conclusions}

Course evaluations indicate that students greatly appreciate the first
advanced course. The use of Haskell as pseudocode does not seem to
confuse them. They can translate it into Racket when asked to do so,
and the Racket code they write on exams does not have Haskell elements
creeping into it. This is probably due to the fact that they never
have to write Haskell, even as pseudocode, during the course. Haskell
intrigues them, and some students express interest in using it. I hope
to develop some optional learning materials for such students in the
near future.

There is more than enough material to fill a first course with topics
approached in a purely functional manner (and one that largely
emphasizes structural recursion). The only real difficulty with
content is the necessity to leave out favourite topics due to the
finite length of the term.

The second advanced course, which needs to move towards mainstream
computer science, is more problematic. The advanced sequence shares
some issues with the major sequence: the more complicated semantics of
mutation; the increased difficulty of testing code written in a
primarily imperative language; the confusing syntax, weak or absent
abstractions, and lack of good support tools associated with popular
languages. Added to these for the advanced sequence are the
disappointment associated with the comparative lack of elegance and
the relatively low-level nature of problem solving typical with
such material. It is not the best advertisement for computer science.

Despite this, students appreciate the second advanced course, perhaps
because all of these elements are present and have even more impact on
students in the second regular course (for majors). They also voice
some of the frustrations that I feel as instructor.  The second course
remains a work in progress, with hope sustained by the fact that
Racket is a good laboratory for language experimentation. With luck I
will soon be able to report on a second course which is as rewarding
for students as the first one.

\section{Bibliography}

\nocite{*}
\bibliographystyle{eptcs}
\bibliography{tfpie}
\end{document}